\documentclass[12pt]{article}
\begin{document}
\title{Quantum Finance: The Finite Dimensional Case
\thanks{Journal of Economic Literature Index Numbers: G10, G12
\hfil\break\indent 2000 Mathematics Subject Classification: 46L53,
91B28\hfil\break\indent This paper was partly written when the
author visited Equipe de Math\'{e}matiques, Universit\'{e} de
Franche-Comt\'{e}, supported by the grant of the academic
cooperation agreement between CAS and CNRS. He would like to thank
the Equipe de Math\'{e}matiques for its kind hospitality. In
particular, he is very grateful to Professor Q.Xu for
conversations on non-commutative probability.}}
\author{Zeqian Chen\\{\small Wuhan Institute of Physics and
Mathematics, Chinese Academy of Sciences}\\
{\small 30 West District, Xiao-Hong Mountain, Wuhan}\\{\small
P.O.Box 71010, Wuhan 430071; China}\\{\small E-mail:
zqchen@wipm.ac.cn}}
\date{}
\maketitle

{\it Abstract}.~~In this paper, we present a quantum version of
some portions of Mathematical Finance, including theory of
arbitrage, asset pricing, and optional decomposition in financial
markets based on finite dimensional quantum probability spaces. As
examples, the quantum model of binomial markets is studied. We
show that this quantum model ceases to pose the paradox which
appears in the classical model of the binomial market.
Furthermore, we re-deduce the Cox-Ross-Rubinstein binomial option
pricing formula by considering multi-period quantum binomial
markets.

\

{\it Key words and phrases}: Selfadjoint operators,
non-commutative stochastic biprocesses, quantum financial markets,
martingale (risk-neutral) states, Pauli spin matrices, the
binomial model, the Cox-Ross-Rubinstein binomial option pricing
formula.

{\it Running Title}: Quantum Finance

\

{\it Contents}

0.~~Introduction

1.~~Notational preliminaries

2.~~Non-commutative martingales and arbitrage-free

3.~~Pricing by no-arbitrage and optimal decomposition

4.~~Quantum binomial models

5.~~Conclusions
\newpage

\section*{0.~Introduction}

Perhaps the most dramatic shift in our understanding of the
physical real-world occurred in 1925, when Heisenberg published a
remarkable paper in which he demonstrated that one could deduce
quantum phenomena from the equations of Newtonian physics provided
one interpreted the time dependent variables as standing for
matrices rather than functions. In contrast to functions, matrices
need not commute under multiplication. Heisenberg's ``matrix
mechanics" quickly attracted the attention of a number of leading
mathematicians, including Jordan, von Neumann, and Weyl. In
particular, von Neumann pointed out that Heisenberg's matrices
were more precisely modelled by self-adjoint Hilbert space
operators. There is now a consensus among scientists that the
classical and relativistic notions of measurement and geometry
that underlie so much of modern mathematics no longer correspond
to our understanding of the real world. Von Neumann was the first
to fully appreciate this fact, and he concluded that we should
seek ``quantized" (= non-commutative) analogues of mathematics. He
proposed that, as in physics, we should begin by replacing
functions by operators [vN55]. During the past sixty years, such
``non-commutative" mathematics have been shown to have a profound
structure theory. For details see for example [Co94], [My93],
[P92], [Ta01], and references therein.

In this paper we try to concern with a more recent innovation, the
{\it quantization of Mathematical Finance.} There are several
reasons why quantizing mathematical finance may be interesting.
First, classical mathematical finance theory is a well established
discipline of applied mathematics (see [EK99] and [Sh99] for
example) which has found numerous applications in financial
markets (see for example [CR85], [Du93], [Hu00], and [Me90]).
Since it is based on probability to a large extend, there is a
fundamental interest in generalizing this theory to the domain of
quantum probabilities. Indeed, recently non-commutative (=
quantum) probability theory has developed considerably. In
particular, all sorts of non-commutative analogues of Brownian
motion and martingales have been studied. We refer to [My93],
[P92], [PX97], and references therein. Second, if the ``Selfish
Genes" [Da76] are reality, we may speculate that stock markets of
human beings are being played already on the molecular level where
quantum mechanics dictates the rules. Third, there is an intimate
connection between the theory of finance and the theory of quantum
games ([My99] and [EWL99]). Indeed, if one buys 100 shares of IBM
stock, he may think that he has made an investment. Ignoring the
broker's fees, he plays a zero-sum game with the seller. If the
price goes up, he will win and the seller will lose; if the price
goes down, he will lose and the seller will win. One of them is
going to profit at the other's expense. It has recently been shown
that quantum strategies can be more successful than classical ones
([My99], [EWL99] and [EM00]). Finally, the quantum version of
financial markets is maybe much more suited to real-world
financial markets rather than the classical one, because the
quantum binomial model ceases to pose the paradox which appears in
the classical model of the binomial market as shown in [Ch2] (see
also \S 4 below).

In retrospect, the field of mathematical finance has undergone a
remarkable development since the seminal papers by F.Black and
M.Scholes [BS73] and R.Merton [Me73], in which the famous
``Black-Scholes Option Pricing Formula" was derived. The idea of
developing a ``formula" for the price of an option actually goes
back as far as 1900, when L.Bachelier wrote a thesis with the
title ``Th\'{e}orie de la sp\'{e}culation" [B00]. It was Bachelier
who firstly had the innovative idea of using a {\it stochastic
process} as a model for the price evolution of a stock. For a
stochastic process $(S_t)_{0 \leq t \leq T}$ he made a natural and
far-reaching choice being the first to give a mathematical
definition of Brownian motion, which in the present context is
interpreted as follows: $S_0$ is today's (known) price of a stock
(say a share of company XYZ to fix ideas) while for the time $t >
0$ the price $S_t$ is a normally distributed random variable.

The basic problem of Bachelier, as well as of modern Mathematical
Finance in general, is that of assigning a price to a contingent
claim. Bachelier used the equilibrium argument. It was the merit
of Black and Scholes [BS73] and Merton [Me73] to have replaced
this argument by a so-called ``no-arbitrage" argument, which is of
central importance to the entire theory. Roughly speaking, an
arbitrage is a riskless way of making a profit with zero net
investment. An economically very reasonable assumption on a
financial market consists of requiring that there are no arbitrage
opportunities. The remarkable fact is that this simple and
primitive ``principle of no arbitrage" allows already to determine
a unique option price in the Black-Scholes model. This is the
theme of the so-called {\it fundamental theorem of asset pricing}
which states briefly that a process $S= (S_t)$ does not allow
arbitrage opportunities if and only if there is an equivalent
probability measure under which $S$ is a martingale.

The history of the fundamental asset pricing theorem goes back to
the seminal work of Harrison, Kreps and Pliska ([HK79], [HP81],
and [K81]). After their pioneering work many authors made
contributions to gradually improve the understanding about this
fundamental theorem, e.g., Duffie and Huang [DH86], Stricker
[St90], Dalang, Morton, and Willinger [DMW90], and [DS94] etc. In
[DS98] this theorem was proved to hold true for very general
(commutative) stochastic processes. More recently, the author
[Ch1] proved a non-commutative version of this theorem.

In \S 3 we deal with this issue in the non-commutative setting based on
finite dimensional Hilbert spaces, after having formalized the
notations of (quantum) arbitrage and quantum trading strategies in
\S 2. The corresponding theorems of pricing by no-arbitrage and optional
decomposition are proved. We also obtain a characterization of complete
markets in the non-commutative setting. Most of our presentation is
inspired by, and follows quite closely, Schachermayer's lecture [Sc01].

For ease of reference a summary of the main results from finite
dimensional quantum probability is given in \S 1 (for details see
[P92]). As examples, the quantum model of binomial markets is
studied in \S 4. We show that this quantum model ceases to pose
the paradox which appears in the classical model of the binomial
market. Furthermore, we re-deduce the Cox-Ross-Rubinstein binomial
option pricing formula by considering multi-period quantum
binomial markets.

\section*{1.~Notational preliminaries}

Throughout this paper, unless otherwise specifically mentioned,
by a Hilbert space ${\bf H}$ we shall always mean a finite dimensional
complex Hilbert spaces with scalar or inner product $<.,.>$
which is conjugate linear in the first and linear in the second variable.
${\bf C}^n$ denotes the $n$-dimensional
complex Hilbert space of all complex $n \times 1$ matrices or column
vectors with the standard inner product$$<u,v> = \sum_j \bar{a}_j b_j$$
where$$u = \left( \begin{array}{l}a_1 \\.\\.\\.\\a_n \end{array}\right),
~v = \left( \begin{array}{l}b_1 \\.\\.\\.\\ b_n \end{array}\right).$$
By the {\it canonical basis} in ${\bf C}^n$ we mean the orthonormal
basis $\{e_1,...,e_n \}$ where $e_j$ is the column vector
with $1$ in the $j$-th position and $0$ elsewhere. When
$n=1$ drop the superscripts and denote the Hilbert space ${\bf C}^1$ by
${\bf C}.$

The set of all operators in ${\bf H}$ is denoted by ${\cal B} ({\bf H}).$
The {\it adjoint} of a operator $A$ is the unique operator $A^*$
satisfying$$< A^* u,v> = <u, A v >$$for all $u,v$ in ${\bf H}.$
${\cal B}({\bf H})$ is an involutive Banach algebra with norm $\|.\|$
and involution $^*.$ Furthermore, for any
$A \in {\cal B}({\bf H}),$ $$\|A \| = \| A^* \| =
\| A^* A \|^{\frac{1}{2}}.$$
In other words ${\cal B}({\bf H})$ is a $C^*$-algebra (indeed, a von
Neumann algebra).

If $\lambda$ is a scalar the same symbol will be frequently used
to denote the operator $\lambda I, I$ denoting identity. For any
$A$ in ${\cal B}({\bf H}),$ $A$ is said to be {\it self-adjoint}
if $A^* = A.$ We write$${\cal O}({\bf H}) = \{ A | A \in {\cal
B}({\bf H}), ~~A = A^* \}$$ and observe that it is a real linear
space. An operator $A$ is said to be {\it positive} if $<u, A u>
\geq 0$ for every $u$ in ${\bf H}.$ A positive operator is
necessarily self-adjoint. If $A_1, A_2$ are in ${\cal O} ({\bf
H})$ we write $A_1 \geq A_2$ if $A_1 - A_2$ is a positive
operator. $\geq$ is a partial order in ${\cal B}({\bf H}).$ By a
{\it projection} we shall always mean an orthogonal projection
onto a subspace of ${\bf H}.$ Denote the set of all projections in
${\bf H}$ by ${\cal P}({\bf H}).$ $A$ is a projection if and only
if $A = A^* = A^2.$ In particular, any projection $E$ is a
positive self-adjoint operator and $0 \leq E \leq 1.$ Thus ${\cal
P}({\bf H}) \subset {\cal O}({\bf H}) \subset {\cal B}({\bf H}).$

For any two elements $u,v$ in ${\bf H}$ we define the operator $|u><v|$ by
$$|u><v| w = <v,w> u$$for all $w$ in ${\bf H}.$ $|u><v|$ is linear in $u$
and conjugate linear in $v$ and satisfies: $(|u><v|)^* = |v><u|$
and $\||u><v| \| = \|u\| \|v\|.$

We shall now describe the quantum analogue of a classical probability space
with $n$ elementary outcomes or sample points. We consider $n$-dimensional
Hilbert space ${\bf H}$ and call any element of
${\cal P}({\bf H})$ an {\it event.} The elements $0$ and $1$ in ${\cal P}
({\bf H})$ are called the {\it null} and {\it certain} events respectively.
If $E_j$ are events we denote by $\cup
E_j$ the event of occurrence of at least one of the $E_j$'s whereas $\cap_j
E_j$ is the event of simultaneous occurrence of all the $E_j$'s, that is,
$\cup_j E_j$ and $\cap_j E_j$ are respectively the projections on
the smallest closed subspace containing the union and joint of
range spaces of all $E_j.$ If $E_1,E_2$ are events and $E_1 \leq E_2$ we
say that $E_1$ implies $E_2.$ If $E$ is an event $1 - E$ is called its
{\it complement.} If $E_1,E_2$ are events then $E_1 + E_2$ is an event
if and
only if $E_1 E_2 = 0.$ Any one dimensional projection $E$
in ${\cal P}({\bf H}
)$ is an {\it atom} in the sense that it cannot be expressed as the sum of
two non-null projections.

For any operator $A$ on the $n$-dimensional Hilbert space ${\bf H}$ and
any orthonormal basis $\{e_1,..., e_n \}$ the quantity
$\sum_j <e_j , A e_j >$
is independent of the basis, called the trace of $A$ and denoted by $tr A.$

A positive operator $\rho$ of unit trace is called a {\it state.}
The set of all states in ${\bf H}$ is denoted by ${\cal S}({\bf
H}).$ For any fixed state $\rho$ the triple $({\bf H}, \rho )$ is
called a {\it simple} or {\it finite dimensional quantum
probability space.} For any $E$ in ${\cal P}({\bf H})$ the
quantity $tr \rho E$ is called the {\it probability of the event
$E$ in the state} $\rho$ and$$tr \rho E = \sum_j < u_j , \rho u_j
>$$where $\{u_j\}$ is an orthonormal basis for the range of $E.$
It follows from the spectral theorem that every state $\rho$ can
be expressed as$$\rho = \sum_j p_j |u_j><u_j|$$where $p_j \geq 0,
\sum p_j = 1$ and $\{u_j \}$ is an orthonormal basis of
eigenvectors of $\rho$ such that$$ \rho u_j = p_j u_j$$for each
$j.$ $\rho$ is said to be {\it faithful} if its eigenvalues are
all greater than zero, that is, $p_j > 0$ for all $j.$ Any one
dimensional projection is called a {\it pure state.} The extreme
points of convex set ${\cal S}({\bf H})$ are precisely the pure
states. In this context it is worth noting that in a sample space
of $n$ elementary outcomes in classical probability the set of all
probability distributions is a convex set whose extreme points are
precisely the $n$ degenerate distributions. In its quantum
analogue the set of pure states is a manifold of dimension $2n
-2.$ It is the richness of the extreme points of convex set ${\cal
S}({\bf H})$ that makes quantum probability worth exploring even
in finite dimensions.

Elements of ${\cal O}({\bf H}),$ i.e., Hermitian operators in
${\bf H},$ are called {\it observables.} An observable in quantum
probability is what a random variable is in classical probability.
Any observable $X,$ being a self-adjoint operator, has the
spectral resolution
$$X = \sum_j x_j E^X_j$$
where $x_1,x_2,...$ are its distinct eigenvalues and $E^X_j$ is the {\it
event that $X$ takes the value $x_j.$} If $g$ is a real valued function
on the real line ${\bf R}$ then
$$g (X) = \sum_j g (x_j) E^X_j$$
is also an observable. The mapping $g \to g (X)$ is a homomorphism
from the algebra of real functions on ${\bf R}$ into the algebra
${\cal B}({\bf H}).$ Events are observables
assuming at most two values $0,1.$

Let $\rho$ be a state and let $X$ be an observable with spectral
resolution $X = \sum_j x_j E^X_j.$ The probability of the event
$E^X_j,$ i.e., $X$ takes the value $x_j$ in the state $\rho,$ is
equal to $tr \rho E^X_j.$ $X$ has {\it expectation} $E_{\rho} (X)$
in the state $\rho$$$E_{\rho} (X) = \sum_j x_j tr \rho E^X_j = tr
\rho \sum_j x_j E^X_j = tr \rho X.$$ For any real valued function
on ${\bf R}$ the expectation $E_{\rho} [g (X)]$ of $g (X)$ in the
state $\rho$ is equal to$$E_{\rho} [g (X)] = \sum_j g(x_j) tr \rho
E^X_j = tr \rho g (X).$$If $u$ is a unit vector in ${\bf H}$ then
in the pure state $u$ (i.e., when $\rho = |u >< u|$) $X$ has the
distribution with mass $<u, E^X_j u>$ at $x_j$ for each $j,$ and
expectation $<u, X u>.$ If $X$ is a non-negative observable or,
equivalently, $X$ is a positive operator then $tr \rho X \geq 0$
for any state $\rho.$ Thus expectation in a state is a
non-negative linear map from ${\cal O}( {\bf H})$ into ${\bf R}$
with value unitary for the observable $1.$

\section*{2.~Non-commutative martingales and arbitrage-free}

In the sequel we shall denote ${\cal B}({\bf H})$ by ${\cal A}$ and assume
that ${\cal A}$ is filtered, so that there exists a
family $( {\cal A}_t )^T_{t =0}$ of unital (closed) $*$-
subalgebras of ${\cal A},$ such that ${\cal A}_s \subset {\cal A}_t$ for
all $s,t$ with $s \leq t,$ and ${\cal A}_0 = {\bf C} I, I$ denoting the
identity on
${\bf H}.$ Given any fixed
state $\rho.$ A sequence $
\{M_t\}^T_{t=0}$ in ${\cal A}$ is said to be a (non-commutative)
martingale with respect to
$({\bf H}, ({\cal A}_t)^T_{t=0}, \rho)$ if it is adapted to
$({\cal A}_t)^T_{t=0}$ and for every $t=1,...,T,$
$$E_{ \rho} A^*  M_t A = E_{ \rho} A^* M_{t-1} A,$$for
all operators $A \in {\cal A}_{t-1}.$

We would like to mention that the non-commutative martingales are
usually defined and studied under a (normal) tracial state (see
[PX97] for example). In that case, the corresponding conditional
expectation operator exists and hence one may define the
martingales as in the classical setting. However, even for a state
$\rho$ in a finite dimensional Hilbert space ${\bf H}$ the
conditional expectation operator $E_{\rho} [.|{\cal B}]$ of a
$*$-subalgebra ${\cal B}$ of ${\cal B}({\bf H})$ need not exist in
general (for details see [Ta72]). Thus we cannot define a
martingale under $\rho$ as in the case of the tracial states or
the commutative setting. Recently, the author [Ch1] generalized
the definition of the non-commutative martingales to the case of
general states as above and show that it is suitable in the
non-commutative generalization of the fundamental theorem of asset
pricing (see also Theorem 2.1 below). In what follows one may find
that this definition is natural and suitable in ``quantum
finance".

Together with $({\cal A}, \rho)$ we shall also consider the
opposite algebra ${\cal A}^{op},$ with the state $\rho^{op},$
namely $\rho = \rho^{op}$ as a linear map on ${\cal A},$ but the
notation is meant to stress the algebra structure we are using.
The spaces ${\cal A}$ and ${\cal A} \otimes {\cal A}$ have natural
${\cal A}-{\cal A}$ bimodule structures given by multiplication on
the right and on the left, namely $A.U.B = AUB$ and $A. ( U
\otimes V). B = A U \otimes V B,$ or equivalently they have a left
${\cal A} \otimes {\cal A}^{op}$-module structure. We shall denote
by $\sharp$ these actions, namely one has $(A \otimes B) \sharp U
= A U B$ and $(A \otimes B) \sharp (U \otimes V) = (A U) \otimes
(V B).$

A ${\cal A}$-valued biprocess is a sequence $H= (H_t)^T_{t=1}$ in the
algebraic tensor product ${\cal A} \otimes {\cal A}^{op}.$
It is called to be
predictable if one has that $H_t \in {\cal A}_{t-1} \otimes
{\cal A}_{t-1}$ for all $t = 1, ...,T.$
In this case, it is clear that one can choose a decomposition
\begin{equation}
H_t = \sum^m_{j=1} A_j \otimes B_j\end{equation} with $A_j, ~B_j
\in {\cal A}_{t-1}$ for $j = 1,...,m$ (in the sequel we shall
always assume that the decompositions we choose satisfy such
properties).

\

{\it Definition 2.1}.~~A {\it quantum model} of a financial market
based on the filtered quantum base $({\bf H}, ({\cal
A}_t)^T_{t=0})$ is an ${\bf R} \times {\cal A}^d$-valued
stochastic process $(B, S),$ where $B = (B_0, B_1,..., B_T)$ is
called a bank account, and
$$S = (S_t)^T_{t=0} = (S^1_t,..., S^d_t )^T_{t=0},$$where $S^j =
(S^j_t)^T_{t=0}$ is a sequence of positive operators with $S^j_t
\in {\cal A}_t,$ which describes the dynamics of the value of the
j-th risk asset $S^j, j=1,..., d.$

\

{\it Definition 2.2}.~~${\cal H}$ denotes the set of {\it quantum
trading strategies} based on the filtered quantum base $({\bf H},
({\cal A}_t)^T_{t=0}).$ An element $H \in {\cal H}$ is an ${\cal
A}$-valued biprocess $(H_t )^T_{t=0}$ such that $H_0 \in {\bf R}$
and for $t= 1,...,T,$
$$H_t = \sum^{m}_{k=1} a_k A^*_k \otimes A_k,$$
where $a_k$ are all real numbers and $A_k \in {\cal A}_{t-1}.$ We
write$${\cal H}^d = \{ (H^1,..., H^d ): H^j \in {\cal H},~j=1,...,
d \}.$$

An {\it investment portfolio} on the $(B, S)$-market is a sequence
$\Pi = (\beta, \gamma)$ where $\beta = (\beta_0, \beta_1, ...,
\beta_T)$ is a sequence of real numbers, and $\gamma = (H^1,...,
H^d) \in {\cal H}^d.$

The {\it value process} of an investment portfolio $\Pi = (\beta,
\gamma)$ on the $(B, S)$-market is the operator sequence $\Pi
\sharp (B, S) = ( [\Pi \sharp (B, S)]_t )^T_{t=0},$ where
$$[\Pi \sharp (B, S)]_t = \beta_t B_t + \sum^d_{j=1} H^j_t \sharp S^j_t.$$

We say that an investment portfolio $\Pi = (\beta, \gamma)$ is
{\it self-financing} if its value $\Pi \sharp (B, S) = ( [\Pi
\sharp (B, S)]_t )^T_{t=0}$ can be represented as following
$$[\Pi \sharp (B, S)]_t = [\Pi \sharp (B, S)]_0 +
\sum^t_{k=1}(\beta_k \triangle B_k + \sum^d_{j=1} H^j_k \sharp
\triangle S^j_k)$$for $t=1,...,T,$ where $\triangle B_k = B_k -
B_{k-1}, \triangle S^j_k = S^j_k - S^j_{k-1}.$

\

{\it Remark 2.1}.~~In what follows, we always assume that
investment portfolios are self-financing. Set
$\overline{B}=(\overline{B}_t)^T_{t=0}$ with $\overline{B}_t = 1$
for all $t=0,1,..., T,$ and $\overline{S}=(
\frac{S_t}{B_t})^T_{t=0},$ which is said to be the discounted
process of $S$ in the $(B, S)$-market. As in the classical
setting, without loss of generality we can replace $(B, S)$ by
$(\overline{B}, \overline{S}).$ In the sequel we shall always
assume that $B_t \equiv 1$ for all $t=0,1,...,T,$ when not
specially stated. In this case, any self-financing investment
portfolio $\Pi = (\beta, \gamma)$ can be reduced to a (d-tuple)
quantum trading strategy $H = (H^1,..., H^d)$ in $S$ so that$$[\Pi
\sharp (B, S)]_t = \alpha + (H \sharp S)_t, ~~t=1,...,T,$$where
$(H \sharp S)_t = \sum^t_{k=1} \sum^d_{j=1} H^j_k \sharp \triangle
S^j_k.$

\

{\bf Lemma 2.1}.~~{\it Let $H$ be a $d$-tuple quantum trading
strategy. If $S = (S^1_t,...,S^d_t)^T_{t = 0}$ is a ${\cal
A}^d$-valued martingale with respect to $({\bf H}, ({\cal
A}_t)^T_{t=0}, \rho)$ then the value process $H \sharp S$ of $H$
in the financial market $S$ is also a martingale with respect to
$({\bf H}, ({\cal A}_t)^T_{t=0}, \rho).$}

\

{\it Proof.}~~Let $H_t = (A^*_1 \otimes A_1,..., A^*_d \otimes
A_d)$ where $A_j \in {\cal A}_{t-1}$ for all $j=1,...,d.$ We have
to prove that for each $m=1,...,T$ $$ E_{ \rho} Y^* [ (H \sharp
S)_m - (H \sharp S)_{m-1}] Y =0$$for all $Y \in {\cal A}_{m-1}.$
One has that$$(H \sharp S)_m - (H \sharp S)_{m-1} = \sum^d_{k=1}
H^k_m \sharp (S^k_m - S^k_{m-1}) = \sum^d_{k=1} A^*_k (S^k_t -
S^k_{t-1} ) A_k$$ when $m=t$ and $= 0$ otherwise. Since $X =
(X_t)_{t \geq 0}$ is a martingale, we get the result. The general
case follows since linear combinations of martingales are
martingales.

\

{\bf Lemma 2.2}.~~{\it Let $S= (S^1_t,...,S^d_t)^T_{t=0}$ be adapted
to $({\bf H}, ({\cal A}_t)^T_{t=0}).$ Then $S$ is a martingale
with respect to $({\bf H}, ({\cal A}_t)^T_{t=0}, \rho)$ if and only if
$$E_{\rho} (H \circ S)_T = 0,$$for every $H \in {\cal H}^d.$}

\

{\it Proof.}~~Suppose that $X = (X_t)_{t \geq 0}$ is a martingale.
By Lemma 2.1 one concludes that$$E_{\rho} (H \sharp S)_T =
E_{\rho}(H \sharp S)_0 =0,$$for each $H \in {\cal H}.$

Conversely, let $Y \in {\cal A}_{t-1}$ for some $t=1,...,T.$ Set
$H^k_t = Y^* \otimes Y.$ Then$$(H \sharp S)_T = Y^* (S^k_t -
S^k_{t-1} )Y,$$and hence $E_{ \rho} Y^* [S^k_t - S^k_{t-1} ] Y
=0.$ This concludes that $(S^k_t)^T_{t=0}$ is a martingale. The
proof is complete.

\

{\it Definition 2.3}.~~We call the subspace ${\cal K}$ of
${\cal O}({\bf H})$ defined by
$${\cal K} = \{ (H \sharp S)_T : H \in {\cal H}^d \}$$
the set of non-commutative contingent claims attainable at price $0.$

\

{\it Remark 2.2}.~~The economic interpretation is the following:
the non-commutative random variables $K = (H \sharp S)_T,$ for
some $H \in {\cal H}^d,$ are precisely those (non-commutative)
contingent claims that an (quantum) economic agent may replicate
with zero initial investment by pursuing some predictable quantum
trading strategy $H.$

For any $\alpha \in {\bf R},$ we call the set of contingent claims
attainable at price $\alpha$ the affine space ${\cal K}_{\alpha}$
obtained by shifting ${\cal K}$ by the constant operator $\alpha,$
in other words the non-commutative random variables of the form
$\alpha + (H \sharp S)_T,$ for some quantum trading strategy $H.$
A quantum financial market $S$ is said to be {\it complete} if
each $A \in O({\cal A}_T) = {\cal A}_T \cap {\cal O}({\bf H})$ is
replicable, that is, $O( {\cal A}_T ) = \cup_{\alpha \in {\bf R}}
{\cal K}_{\alpha}.$

\

{\it Definition 2.4}.~~We call the convex cone ${\cal C}$ in
$O( {\cal A}_T )$ defined by
$${\cal C} = \{ C \in O({\cal A}_T): ~there~is~some~K \in {\cal K},
K \geq C \}$$
the set of non-commutative contingent claims super-replicable at
price $0.$

\

Observe that ${\cal C}$ is a convex cone containing the negative elements
$\{A \in O({\cal A}_T): A \leq 0 \}.$

Economically speaking, a non-commutative contingent claim $A \in
O({\cal A}_T)$ is super-replicable at price $0,$ if one quantum
agent can achieve it with zero net investment, subsequently
pursuing some predictable quantum trading strategy $H$--thus
arriving at some non-commutative contingent claim $K$--and then,
possibly, ``throwing away money" to arrive at $A.$ This operation
of ``throwing away money" may seem awkward at this stage, but we
shall see later that the set ${\cal C}$ plays an important role in
the development of the present theory, as in the commutative
setting.

\

{\it Definition 2.5}.~~A quantum financial market $S$ satisfies the
no-arbitrage condition (NA) if$${\cal K} \cap {\cal A}_+ = \{ 0\}$$or,
equivalently,
$${\cal C} \cap {\cal A}_+ = \{ 0\},$$where ${\cal A}_+ =
\{ A \in {\cal A}: A \geq 0 \}.$

\

In other words we now formalize the concept of an (quantum)
arbitrage possibility: it consists of the existence of a $d$-tuple
quantum trading strategy $H$ such that---starting from an initial
investment zero---the resulting contingent claim $f = (H \sharp
S)_T$ is non-negative and not identically equal to zero. If a
(quantum) financial market does not allow for arbitrage we say
that it satisfies the {\it no-arbitrage condition} (NA).

\

{\it Definition 2.6}.~~A state $\rho$ on ${\bf H}$ is called a
martingale state of $S = (S^1_t,...,S^d_t)^T_{t = 0},$ if $S$ is a
${\cal A}^d$-valued martingale with respect to $({\bf H}, ({\cal
A}_t)^T_{t=0}, \rho).$

We denote by $M(S)$ (or, $M_f(S)$) the family of all such
(faithful) martingale states, and say that $S$ satisfies the
condition of the existence of a faithful martingale state (EMS) if
$M_f(S) \not= \emptyset.$

As usual, a faithful martingale state is called a {\it
risk-neutral state} of the market.

\

{\bf Lemma 2.3}.~~{\it For a state $\rho$ on ${\bf H}$ the following are
equivalent:

$(1)$~~$\rho \in M(S),$

$(2)$~~$E_{\rho} [K] = 0,$ for all $K \in {\cal K},$

$(3)$~~$E_{\rho} [C] \leq 0,$ for all $C \in {\cal C}.$}

\

{\it Proof.}~~The equivalences are rather trivial, the equivalence of
(1) and (2) immediately follows from Lemma 2.2 while the equivalence of
(2) and (3) is straightforward.

\

After having fixed these formalities we may formulate and prove a
quantum analogue of the central result of the finance theory of
pricing and hedging by no-arbitrage, the so-called fundamental
theorem of asset pricing, which goes back to Harrison and Pliska
[HP81] in the classical case.

\

{\bf Theorem 2.1}~~{\it For a quantum financial market $S$
modelled on a finite dimensional quantum stochastic base $({\bf
H}, ({\cal A}_t)^T_{t=0})$ it satisfies $($NA$)$ if and only if
$M_f (S) \not= \emptyset.$}

\

{\it Proof}.~~(EMS) $\rightarrow$ (NA): By Lemma 2.2 we have that
$E_{\sigma}[C] \leq 0$ for each $\sigma \in M_f (S)$ and $C \in {\cal C}.$
However, if (EMS) would hold and (NA) were violated, there would exist a
$\sigma \in M_f(S)$ and $C \in {\cal C}, C > 0,$ whence $E_{\sigma} [C]
> 0$ since $\sigma$ is faithful, a contradiction.

(NA) $\rightarrow$ (EMS): Since ${\cal S}({\bf H})$ is a convex,
compact subset
of ${\cal O}({\bf H})$ and, by the (NA)  assumption, disjoint from
${\cal K},$ there is $Q \in {\cal O}({\bf H})$ and $\alpha  < \beta$ such
that$$tr [K Q] \leq \alpha,~~K \in {\cal K};$$ $$tr  [V Q]  \geq \beta ,
~~ V \in {\cal S}({\bf H}).$$
As ${\cal K}$ is a linear space, we have that $\alpha \geq 0.$
Hence $ \beta > 0.$
Therefore $A \to tr [QA]$ is a positive, faithful linear functional
on ${\cal O}({\bf H}).$ Normalize $Q$ we obtain a faithful martingale
state of $S$ by Lemma 2.3.

\

{\it Remark 2.3}.~~Theorem 2.1 is a special case of the
non-commutative version of the fundamental asset pricing theorem,
which was proved in [Ch1]. However, the proof presented here is
different from that of [Ch1].

\

{\bf Theorem 2.2}.~~{\it Let $S$ satisfy $(NA)$ and $A \in O({\cal
A}_T)$ so that\begin{equation}A = \alpha + (H \sharp
S)_T,\end{equation} for some $\alpha \in {\bf R}$ and some trading
strategy $H.$ Then, the constant $\alpha$ is uniquely determined
by $($2$)$ and
\begin{equation}\alpha = E_{\sigma} [ A]\end{equation}}
for every $\sigma \in M_f (S).$

\

{\it Proof}.~~Suppose that there were two representations $A =
\alpha_1 + (H^1 \sharp S)_T$ and $A = \alpha_2 + (H^2 \sharp S)_T$
with $\alpha_1 \not= \alpha_2.$ Assuming $\alpha_1 > \alpha_2$ we
find an obvious arbitrage possibility: we have$$\alpha_1 -
\alpha_2 = ( [H^1 - H^2] \sharp S)_T,$$ that is, the trading
strategy $H^1 - H^2$ produces a strictly positive result at time
$T,$ a contradiction to (NA).

The equation (3) results from the fact that, for every quantum
trading strategy $H$ and every $\sigma \in M_f (S),$ the process
$(H \sharp S)$ is a martingale under $\sigma.$ The proof is
complete.

\

{\it Remark 2.4}.~~Theorem 2.2 says that if a contingent claim is
replicable, its price is just the expectation of its payoff with
respect to a (or, any) faithful martingale state on the underlying
market. Moreover, the converse to Theorem 2.2 still holds true,
see Theorem 3.1 below.

\section*{3.~Pricing by no-arbitrage and optimal decomposition}

Denote by $cone[ M(S) ]$ and $cone[ M_f (S) ]$ the cones generated by
the convex sets $M(S)$ and $M_f (S)$ respectively. As following we shall
clarify the polar relation between these cones and the cone ${\cal C}.$

Recall that, for a pair $(E, E^*)$ of vector spaces in separating duality
via the scalar product $<.,.>,$ the polar $Q^0$ of a set $Q$ in $E$ is
defined
as$$Q^0 = \{ g \in E^* : < f, g> \leq 1, f \in Q \}.$$
In the case when $Q = {\cal C}$ which is a closed convex cone we have that
$$
{\cal C}^0 = \{ g \in E^* : < C, g> \leq 0, ~for~all~C \in {\cal C} \}.$$

The bipolar theorem (see for example [Sch66]) states that the bipolar
$Q^{00}:
= (Q^0)^0$ of a set $Q$ in $E$ is the $\sigma (E, E^*)$-closed convex
hull of
$E.$ Note that in our finite dimensional setting ${\cal C}$ is closed.
Hence
we deduce from the bipolar theorem that ${\cal C} = {\cal C}^{00}.$

\

{\bf Lemma 3.1}.~~{\it Suppose that $S$ satisfies $(NA).$ Then the
polar of ${\cal C}$ is equal to $cone[ M(S) ]$ and $M_f (S)$ is
dense in $M(S).$ Hence the following assertions are equivalent
for $A \in O( {\cal A}_T ):$

$(a)$~~~$A \in {\cal C};$

$(b)$~~~$E_{\sigma} [ A] \leq 0,$ for all $\sigma \in M(S);$

$(c)$~~~$E_{\sigma} [ A] \leq 0,$ for all $\sigma \in M_f (S).$}

\

{\it Proof}.~~The fact that the polar ${\cal C}^0$ and $cone[M(S)]$
coincide,
follows from Lemma 2.3. Hence the equivalence of (a) and (b) follows
from the bipolar theorem.

As regards the density of $M_f (S)$ in $M(S)$ we first deduce
from Theorem 2.1
that there is at least one $\rho \in M_f (S).$ For any $\sigma \in
M(S)$ and $0 < \alpha \leq 1$ we have that $\alpha \rho + (1 - \alpha )
\sigma \in M_f (S),$ which clearly implies that $M_f (S)$ is dense in
$M(S).$ The equivalence of (b) and (c) is obvious.

\

For an element $A \in O( {\cal A}_T ),$ we call
$\alpha \in {\bf R}$ an {\it arbitrage-free price,}
if$${\cal C}^{A, \alpha} \cap {\cal A}_+ = \{ 0 \},$$
where ${\cal C}^{A, \alpha}$ denotes the cone spanned by ${\cal C}$ and
the linear space spanned by $A - \alpha.$

The next theorem tells us precisely what the quantum principle of
no-arbitrage can tell us about the possible prices for a non-commutative
contingent claim $A.$ In the classical case it goes back to the work
of D.Kreps [K81].

\

{\bf Theorem 3.1}~~{\it Assume that $S$ satisfies $(NA)$ and $A \in
O( {\cal A}_T ).$ Define\begin{equation}\overline{\pi} (A) =
\sup \{ E_{\sigma} [A] : \sigma \in M_f (S) \},\end{equation}and
\begin{equation}\underline{\pi} (A)
= \inf \{ E_{\sigma} [A] : \sigma \in M_f (S) \}.\end{equation}

Either $\underline{\pi} (A) = \overline{\pi} (A),$ in which case $A$ is
attainable at price $\pi (A): = \underline{\pi} (A) = \overline{\pi} (A),$
i.e., $A = \pi (A) + (H\circ S)_T$ for some $H \in {\cal H};$ therefore $
\pi (A)$ is the unique arbitrage-free price for $A.$

Or $\underline{\pi} (A) < \overline{\pi} (A),$ in which case $\{ E_{\sigma}
[A]: \sigma \in M_f (S) \}$ equals the open interval $
(\underline{\pi} (A) , \overline{\pi} (A)),$ which in turn equals the set
of arbitrage-free prices for the non-commutative contingent claim $A.$}

\

{\it Proof}.~~First observe that the set $\{ E_{\sigma} [A]: \sigma \in
M_f (S) \}$ forms a bounded nonempty interval in ${\bf R},$ which we denote
by $I.$ We claim that a number $\alpha \in I$ if and only if $\alpha$ is an
arbitrage-free price for $A.$ Indeed, supposing that $\alpha \in I$
we may find $\sigma \in M_f (S)$ such that $E_{\sigma} [A] = \alpha$
and hence, ${\cal C}^{A,
\alpha} \cap {\cal A}_+ = \{ 0\}.$

Conversely, suppose that ${\cal C}^{A, \alpha} \cap {\cal A}_+ = \{ 0\}.$
Note that ${\cal C}^{A, \alpha}$ is a closed convex cone,
by the same argument as in the proof
of Theorem 2.1 one concludes that there exists a faithful state
$\sigma$ such that $\sigma [R] \leq 0$ for all $R \in
{\cal C}^{A, \alpha}.$ This
implies that $E_{\sigma} [A] = \alpha,$ that is, $\alpha \in I.$
Suppose that $\alpha = \overline{\pi}(A),$ and consider
$A - \overline{\pi}(A)
.$ By definition we have that $E_{\sigma} [A - \overline{\pi}(A) ] \leq 0,$
for all $\sigma \in M_f (S),$ and therefore by Lemma 3.1, that $A -
\overline{\pi}(A) \in {\cal C}.$ We may find $K \in {\cal K}$ such that
$K \geq A -
\overline{\pi}(A).$ If the sup in (4) is attained, that is, if there is
$\rho \in M_f (S)$ such that $E_{\rho} [A] = \overline{\pi}(A),$
then we have that$$0 = E_{\rho} [K] \geq E_{\rho} [A - \overline{\pi}
(A) ] =0$$ which implies that $K = A - \overline{\pi}(A);$ in other
words $A$ is
attainable at price $\overline{\pi}(A).$ This in turn implies that
$E_{\sigma}
[A] = \overline{\pi}(A),$ for all $\sigma \in M_f (S),$
and thus $I$ is reduced to the singleton $\{\overline{\pi}(A) \}.$

Hence, if $\underline{\pi}(A) < \overline{\pi}(A),\overline{\pi}(A)$ cannot
belong to the interval $I,$ which is therefore open on the right hand side.
Passing from $A$ to $- A$ we obtain the analogous result for the left hand
side of $I,$ which thus concludes that $I = (\underline{\pi}(A),
\overline{\pi}(A) ).$ The proof is complete.

\

{\bf Corollary 3.1}~~{\it For a quantum
financial market $S$ satisfying $(NA)$ the following are equivalent:

$(i)$~~$M_f (S)$ consists of a single element $\rho$ restricted
on ${\cal A}_T,$ in the sense that $\rho = \sigma$ if and only if
$E_{\rho} A = E_{\sigma} A$ for all $A \in {\cal A}_T.$

$(ii)$~Each element $A \in O( {\cal A}_T )$ may be represented as
\begin{equation}A = \alpha + (H \sharp S)_T,\end{equation}
for some $\alpha \in {\bf R}$ and some trading strategy $H.$ In this case
$\alpha = E_{\rho} [A].$}

\

{\it Proof}.~~The implication (i) $\rightarrow$ (ii) immediately follows
from Theorem 3.1. For the implication (ii) $\rightarrow$ (i), note that (6)
implies that $\alpha = E_{\rho} [A]$ for all $\rho \in M_f (S).$ Hence,
if $M_f (S)$ contains
two different elements $\rho_1$ and $\rho_2$ restricted
on ${\cal A}_T,$ we may find that an
element $A \in O( {\cal A}_T )$ so that $E_{\rho_1} [A] \not=
E_{\rho_2} [A].$ This completes the proof.

\

Corollary 3.1 is the non-commutative analogue of the
``second fundamental asset pricing theorem" as called in [Sh99].
It shows that an arbitrage-free (quantum) financial market $S$ is complete
if and only if $M_f (S) = \{ \rho \}$ for some faithful state $\rho$
on ${\bf H},$ in the sense that $\rho = \sigma$ if and only if
$E_{\rho} A = E_{\sigma} A$ for all $A \in {\cal A}_T.$

As following is a dynamic version of Theorem 3.1 on pricing by
no-arbitrage, which holds true in a general commutative setting
(see [K96]).

\

{\bf Theorem 3.2}~~{\it Suppose that $S$ satisfies $(NA)$ and let
$V = (V_t)^T_{t=0}$ be an adapted self-adjoint stochastic process.
The following assertions are equivalent:

$(a)$~~$V$ is a super-martingale for each $\rho \in M_f (S)$ in
the sense that, for every $t=1,...,T,$$$ E_{\rho} A^* V_t A \leq
E_{\rho} A^* V_{t-1} A,$$for all $A \in {\cal A}_{t-1}.$

$(b)$~~$V$ can be decomposed into$$V = V_0 + H \sharp S - C,$$
where $H \in {\cal H}$ and $C = (C_t )^T_{t=0}$ is an increasing
adapted process with starting at $0,$ that is, $0 = C_0 \leq
C_{t-1} \leq C_t$ for all $t=1,..., T.$}

\

{\it Proof}.~~First assume that $T=1,$ i.e., we have a one-period
model $S = (S_0, S_1).$ Since ${\cal A}_0$ is trivial, $V_0 =$ some
$\alpha \in {\bf R}.$ Assuming (a) we concludes from Lemma 3.1 that
there is a quantum trading strategy $H$ such that
$$V_0 + (H \sharp S)_1 \geq V_1.$$
Letting $C_0 =0$ and writing $\Delta C_1 = C_1 = V_0 + (H \sharp
S)_1 - V_1$ we obtain the required decomposition.

Note that Lemma 3.1 holds true without assumption that ${\cal A}_0 =
{\bf C} I.$ We apply the above argument
to the one-period financial market $(S_{t-1}, S_t)$ adapted to the
filtration $\{ {\cal A}_{t-1}, {\cal A}_t\}.$ We thus obtain a
$H_t \in {\cal A}_{t-1}$ such that
$$V_{t-1} + H_t \sharp \Delta S_t \geq V_t.$$
Setting $\Delta C_t = V_{t-1} + H_t \sharp \Delta S_t - V_t$
yields that
$$\Delta V_t = H_t \sharp \Delta S_t - \Delta C_t.$$
This finishes the construction of the optional decomposition:
define the predictable process $H$ as $(H_t)^T_{t=1},$ and the
adapted increasing process $C$ by $C_t = \sum^t_{j=1} \Delta C_j.$
This shows that (a) implies (b).

The converse implication is trivial. The proof is complete.

\

A process of the form $V = V_0 + H \sharp S - C$ can be though of
the wealth process of an economic (quantum) agent, starting at an
initial wealth $V_0,$ subsequently investing in the quantum
financial market according to the quantum trading strategy $H,$
and consuming as described by the process $C:$ the random variable
$C_t$ models the accumulated consumption during the time interval
$\{1,...,T\}.$ The above theorem states economically that these
wealth processes are characterized by condition (a).

\section*{4.~Quantum binomial models}

{\bf Example 1}. (A single-step model)~~We consider a simple
`single-step' model of a $(B, S)$-market formed by a bank account
$B = (B_0, B_1)$ and some stock of price $S = (S_0, S_1).$ We
assume that the constants $B_0$ and $S_0$ are positive and
$$B_1 = B_0 (1 + r),~~S_1 = S_0 ( 1 + A),$$
where the interest rate $r$ is a constant ($r > -1$) and
the rate $A$ is an observable with the spectral resolution
$$A = \sum^m_{j=1} a_j E^A_j,$$
where $a_j > -1$ for all $j=1,...,m.$

Along with the $(B,S)$-market we can consider its discounted
market $(\bar{B}, \bar{S}),$ where
\begin{center}$\bar{B} = (\bar{B}_0, \bar{B}_1)$ with $\bar{B}_0 =
\bar{B}_1 = 1$\end{center}and
\begin{center}$\bar{S} = (\bar{S}_0, \bar{S}_1)$ with $\bar{S}_0 =
\frac{S_0}{B_0},~~\bar{S}_1 = \frac{S_1}{B_1}$.\end{center}Then,
it is easy to check that for $\rho \in {\cal S} ({\bf H}),
E_{\rho} \bar{S}_1 = \bar{S}_0$ if and only if
\begin{equation}E_{\rho} A = r.\end{equation}

In the sequel, we let ${\bf H} = {\bf C}^2$ with its canonical
basis $|0> = \left(
\begin{array}{l}1 \\  0 \end{array}\right), |1> = \left(
\begin{array}{l} 0 \\ 1\end{array} \right).$ Then ${\cal O}({\bf
H})$ has the basis
$$I_2 =\left( \begin{array}{ll}1 &  0 \\  0 & 1\end{array}
\right), ~\sigma_x =\left( \begin{array}{ll}0 &  1 \\  1 &
0\end{array} \right), ~\sigma_y =\left( \begin{array}{ll}0 &  -i
\\  i & ~0\end{array} \right),~\sigma_z =\left( \begin{array}{ll}1
& ~0 \\  0 & -1\end{array} \right),$$where $\sigma_x, \sigma_y,$
and $\sigma_z$ are the well-known Pauli spin matrices of quantum
mechanics. Set
$$A = x_0 I_2 + x_1 \sigma_x + x_2 \sigma_y + x_3 \sigma_z,$$
which takes two values$$a = x_0 - \sqrt{x^2_1 + x^2_2 + x^2_3},~~b
= x_0 + \sqrt{x^2_1 + x^2_2 + x^2_3},$$where $x_j$ are all real
numbers such that $x^2_1 + x^2_2 + x^2_3 \neq 0$ and $ a, b > -1.$

\

{\bf Theorem 4.1}~~{\it With notations as above, the single-period
quantum binomial model is arbitrage-free if and only if $-1 < a <
r < b.$ In this case, the set of its risk-neutral states consists
of\begin{equation}\label{e8}\rho = \frac{1}{2} (I_2 + x \sigma_x +
y \sigma_ y + z \sigma_z ) = \frac{1}{2}\left(
\begin{array}{ll}1 + z &  x- iy  \\  x+ iy & 1
- z\end{array} \right),\end{equation}where all $(x,y,z)$
satisfy\begin{equation}\label{e9}\left\{ \begin{array}{l} x^2 +
y^2 + z^2 < 1,\\x_1 x + x_2 y + x_3 z = r - \frac{a+b}{2},
\end{array} \right. \end{equation}which is a disk of radius
$\sqrt{1- \frac{(2 r - a-b)^2}{(b-a)^2}}$ in the unit ball of
${\bf R}^3.$}

\

{\it Proof}.~~Given$$\rho = \frac{1}{2} (w I_2 + x \sigma_x + y
\sigma_ y + z \sigma_z ) = \frac{1}{2} \left( \begin{array}{ll}w +
z &  x- iy \\  x+ iy & w - z\end{array} \right),$$which takes two
values
$$\lambda_1 = \frac{1}{2} \left(w - \sqrt{x^2 + y^2 + z^2} \right),
~~\lambda_2 = \frac{1}{2} \left( w + \sqrt{x^2 + y^2 + z^2}
\right).$$Then, $\rho$ is a faithful state if and only if $t r
\rho =1$ and $\lambda_1 > 0.$ This concludes that $w = 1$ and$$x^2
+ y^2 + z^2 < 1.$$Moreover, by $t r \rho A =r$ one concludes
that$$x_1 x + x_2 y + x_3 z = r - \frac{a+b}{2},$$which completes
the proof.

\

We are interested in European call options in the single-period
quantum binomial market introduced above. Let $K$ be the exercise
price of the European call option. Then its payoff is of form$$H =
(S_1 - K)^+,$$which takes values$$h_a = [ S_0 (1 + a) - K ]^+$$and
$$h_b = [ S_0 (1 + b) - K ]^+.$$

Let $C$ be the option value at time $0.$ Since ${\cal A}_0 = {\bf
C} I,$ there is a replicating portfolio of a pair $(\beta,
\gamma)$ of real numbers such that
$$H = \beta B_1 + \gamma S_1 = \beta B_0 + \gamma S_0 +
\beta \triangle B_1  + \gamma \triangle S_1.$$A simple computation
yields that$$\beta = \frac{1}{B_0 (1+r)} \frac{h_a S_b - h_b
S_a}{S_b - S_a},~~\gamma = \frac{h_b - h_a}{S_b - S_a},$$where
$S_a = S_0 ( 1 + a ), S_b = S_0 (1 + b).$ Hence $H$ is replicable
and, by Theorem 2.2,\begin{equation}\label{e10}C = \beta B_0 +
\gamma S_0 = \frac{1}{1+ r} \left( \frac{b-r}{b-a} h_a +
\frac{r-a}{b-a} h_b \right ).\end{equation}

Also, it is easy to check that
\begin{equation}\label{e11}C = E_{\rho} H = \frac{1}{1+ r}
\left ( \frac{b-r}{b-a} h_a + \frac{r-a}{b-a} h_b \right
)\end{equation}for all risk-neutral states $\rho$ in (8) and (9),
as stated in (3) of Theorem 2.2.

Recall that the classical single-period binomial model is formed
by a bank account $B = (B_0, B_1)$ and some stock of price $S =
(S_0, S_1)$, in which one assumes that the constants $B_0$ and
$S_0$ are positive,$$B_1 = B_0 (1 + r),~~S_1 = S_0 ( 1 +
R),$$where $R$ is a random variable taking just two values, $a$
and $b,$ such that$$-1 < a < r < b.$$It is well known that the
European call option pricing formula on this model is the same as
equation (10).

It is surprising and seems counterintuitive that the option
pricing formula in equation (10) does not involve the
probabilities of the stock price moving up ($R= b$) or down ($R =
a$) in the classical case. For example, we get the same option
price when the probability of $R= b$ is $0.5$ as we do when it is
$0.9.$ However, a classical view, dating back to the times of
J.Bernoulli and C.Huygens, is that$$\frac{1}{1 + r}E(S_1 - K)^+ =
\frac{1}{1 + r }\left( p h_b + (1-p) h_a \right )$$with $p ={\bf
P}(R = b),$ could be a reasonable price of such an option in the
classical random model. It should be emphasized, however, that
this quantity depends essentially on our assumption on the value
$p.$ This is not the case.

There are some explanations on this paradox in the classical model
of the binomial market, see for example [Hu00, p.205]. However,
the quantum model ceases to pose the paradox. The key reason that
we conclude from the single-period quantum binomial market
introduced above is that there do not exist at all the
probabilities of the stock price moving up or down in the
single-period quantum binomial market. In the single-period
quantum binomial market all individuals are indifferent to the
movement in the stock price. The probability of an upward movement
in the stock price appearing in the classical case comes from one
observer's measurement to the stock. Each individual is indeed in
favor of some different value of its upward or down probability.
Precisely, suppose that$$A = a |u><u| + b |v><v|,$$where $\{u, v
\}$ is an orthonormal basis of ${\bf C}^2.$ Then, for every state
$\rho \in {\cal S} ( {\bf C}^2 )$ the probabilities of that $A$
takes the values $a$ and $b$ in the state $\rho$ are equal to$$tr
\rho |u><u| = <u , \rho u
>$$and$$tr \rho |v><v| = < v, \rho v
>,$$ respectively, which are those probabilities appearing in the
classical setting when one observer measures the stock market on
the state $\rho.$ (For more details see [Ch2].) From this one
could conclude that the quantum binomial market is much more
suited to real-world financial markets rather than the classical
one.

\

{\bf Example 2}. (The Cox-Ross-Rubinstein binomial option pricing
formula)~~Since there is a paradox in the classical model of the
binomial market, the classical random model is not suitable for
getting the Cox-Ross-Rubinstein binomial option pricing formula.
As following we shall re-deduce this famous formula by using
multi-period quantum binomial markets.

Let ${\bf H}_n = ( {\bf C}^2 )^{\otimes n}$ and write$$
|\varepsilon_1 ... \varepsilon_n> = |\varepsilon_1
> \otimes ... \otimes |\varepsilon_n>,~~\varepsilon_1 ,...,
\varepsilon_n = 0,1.$$ Then, $\{ |\varepsilon_1 ... \varepsilon_n>
: \varepsilon_1 ,..., \varepsilon_n = 0,1 \}$ is the canonical
basis of ${\bf H}_n.$ Given $-1 < a < r < b,$ we define a
$N$-period quantum binomial market $(B,S)$ with $B=(B_0,
B_1,...,B_N)$ and $S=(S_0, S_1,...,S_N)$ as following:
\begin{equation}\label{e13}B_n = B_0 (1 + r)^n,~~S_n = S_0
\bigotimes^n_{j=1} (1 + A_j) \otimes I_{ N -
n},~~n=1,...,N,\end{equation} where the constants $B_0$ and $S_0$
are positive, $I_{N-n}$ is the identity on ${\bf H}_{N-n}$ and,
$$A_j = \frac{a+b}{2} I_2 + x_{1j} \sigma_x + x_{2j} \sigma_y + x_{3j}
\sigma_z,$$where$$x^2_{1j} + x^2_{2j} + x^2_{3j} =
\frac{(b-a)^2}{4}$$for all $j=1,...,N.$

Consider European call options in the $N$-period quantum binomial
market $(B, S).$ Its payoff is $$H_N = (S_N - K)^+,$$where $K$ is
the exercise price of the European call option. There are $N$
orthonormal bases $\{ (u_j, v_j ),j=1,...,N \}$ in ${\bf C}^2$
such that$$\begin{array}{lcl}S_N & =& S_0 \bigotimes^N_{j=1} (1 +
A_j )\\~\\&=& S_0 \bigotimes^N_{j=1} [ (1 + b) |u_j >< u_j | + (1
+ a) |v_j >< v_j | ]\\~\\& = & S_0 \sum^N_{n=0} (1 + b)^n (1 +
a)^{N-n} [ \sum_{| \sigma | = n} \bigotimes^N_{j=1} | w_{j \sigma}
>< w_{j \sigma} | ]\end{array}$$where all $\sigma$ are subsets of
$\{ 1,..., N \}, w_{j \sigma} = u_j$ for $j \in \sigma$ or $w_{j
\sigma} = v_j$ otherwise. Then$$(S_N - K)^+ = \sum^N_{n=0} [ S_0
(1 + b)^n (1 + a)^{N-n} - K ]^+ [ \sum_{| \sigma | = n}
\bigotimes^N_{j=1} | w_{j \sigma}
>< w_{j \sigma} | ],$$which is replicable in $(B, S)$ by
induction.

By Theorem 4.1 one has that all states of the form
\begin{equation}\label{e15}\bigotimes^N_{j=1} \rho_j =
\frac{1}{2^N} \bigotimes^N_{j=1} ( I_2 + x_j \sigma_x + y_j
\sigma_ y + z_j \sigma_z)\end{equation}are faithful martingale
states of the $N$-period quantum binomial market $(B, S),$ where
$(x_j, y_j, z_j)$ satisfies$$\left\{ \begin{array}{l} x^2 + y^2 +
z^2 < 1,\\x_{1j} x + x_{2j} y + x_{3j} z = r - \frac{a+b}{2},
\end{array} \right.$$for every $j=1,...,N.$ Hence, by the
principle of risk-neutral valuation, the price $C^N_0$ at time $0$
of the European call option $(S_N - K)^+$ is given
by$$\begin{array}{lcl}C^N_0 & = & tr [ ( \bigotimes^N_{j=1} \rho_j
) (S_N - K)^+ ]\\~\\& = & (1 + r)^{-N} \sum^N_{n=0} \frac{N !}{n !
(N - n) !} q^n (1-q)^{N - n} [ S_0 (1 + b)^n (1 + a)^{N-n} - K
]^+\\~\\& = & S_0 \sum^N_{n = \tau} \frac{N !}{n ! (N - n) !} q^n
(1-q)^{N - n} \frac{(1 + b)^n (1 + a)^{N-n}}{(1 + r)^N}\\~\\&~& -
K (1 + r)^{-N} \sum^N_{n = \tau} \frac{N !}{n ! (N - n) !} q^n
(1-q)^{N - n},\end{array}$$where $q = \frac{r - a}{b-a},$ and
$\tau$ is the first integer $n$ for which$$S_0 (1+ b)^n (1 +
a)^{N-n} > K.$$

Now observe that using $q = \frac{r - a}{b-a}$ and$$q^{\prime} = q
\frac{1 + b}{1 + r}$$we obtain $0 < q^{\prime} < 1$ so that we can
finally write the fair price for the European call option in this
multi-period quantum binomial pricing model
as\begin{equation}\label{e16}C^N_0 = S_0 \Psi (\tau ; N,
q^{\prime} ) - K (1 + r)^{-N} \Psi (\tau ; N,
q),\end{equation}which is the well-known {\it Cox-Ross-Rubinstein
binomial option pricing formula} for the European call. Here
$\Psi$ is the complementary binomial distribution function, that
is,$$\Psi (m; n, p) = \sum^n_{j=m} \frac{n !}{j ! (n-j) !} p^j (
1-p)^{n-j}.$$

We would like to point out that the $N$-period quantum binomial
market can be physically realized by the system of $N$
distinguishable particles whose statistics are described by $( (
{\bf C}^2 )^{\otimes N}, \{ {\cal A}_n \}^N_{n=0} ),$ where ${\cal
A}_0 = {\bf C} I_N,$ and$${\cal A}_n = {\cal B} ( ( {\bf C}^2
)^{\otimes n} ) \otimes I_{N-n} = \{ B \otimes I_{N-n} : B \in
{\cal B} ( ( {\bf C}^2 )^{\otimes n} ) \}$$for $n=1,..., N.$ We
will discuss this naturally physical realization of the $N$-period
quantum binomial market in a forthcoming paper.

\section*{5.~Conclusions}

We transfer methods of quantum mechanics to concepts used in
Mathematical Finance. The quantum theory for arbitrage, asset
pricing, and optional decomposition in financial markets based on
finite dimensional quantum probability spaces is presented. To
reveal what features of real-world financial markets are captured
by the quantum model of the financial markets, we have studied the
quantum model of the binomial markets. It was shown that the
quantum binomial model could cease to pose the paradox which
appears when pricing an option in the classical random model of
the binomial markets. From this one could conclude that the
quantum model of a financial market is much more suited to
real-world financial markets rather than the classical one.
Moreover, since the classical random model is not appropriate for
the binomial markets, we re-deduce the Cox-Ross-Rubinstein
binomial option pricing formula by using multi-period quantum
binomial markets.

\section*{Acknowledgment}

The author is very grateful to Professor Dr. D.Sondermann for
useful advice and comments. Some parts of \S 4 were at his
suggestion.

\begin{center}R{\small EFERENCES}\end{center}
\begin{description}
\item[B00~~]~L.Bachelier, Th\'{e}orie de la sp\'{e}culation, {\it Ann. Sci.
\'{E}cole Norm. Sup.,} 17, 21-86 (1900). English translation in:
{\it The Random Character of Stock Market Prices,} P.Cootner ed.,
Cambridge, Mass.: MIT Press, 1964, pp 17-78
\item[BS73~]~F.Black, M.Scholes, The pricing of options and corporate
liabilities, {\it J. Political Econ.,} 81, 637-659 (1973)
\item[Ch1~~]~Zeqian Chen, A non-commutative version of the
fundamental theorem of asset pricing, {\it Preprint}
(www.arxiv.org/quant-ph/0112159)
\item[Ch2~~]~Zeqian Chen, Quantum theory for the binomial model in
finance theory, {\it Preprint} (www.arxiv.org/quant-ph/0112156v5)
\item[Co94~]~A.Connes, {\it Noncommutative geometry,} San Diego: Academic
Press, Inc., 1994
\item[CRR79]~J.C.Cox, S.A.Ross, M.Rubinstein, Option pricing: A
simplified approach, {\it J. Finance Econ.,} 7, 229-263 (1979)
\item[CR85~]~J.C.Cox, M.Rubinstein, {\it Options markets,} Prentice Hall:
Prentice-Hall, Inc., 1985
\item[DMW90]~R.C.Dalang, A.Morton, W.Willinger, Equivalent martingale
measures and no-arbitrage in stochastic securities market model,
{\it Stochastics Stoch. Rep.,} 29, 185-201 (1990)
\item[Da76~~]~R.Dawkins, {\it The Selfish Gene}, Oxford: Oxford
University Press, 1976
\item[DS94]~F.Delbaen, W.Schachermayer, A general version of
the fundamental theorem of asset pricing, {\it Math. Ann.,} 300,
463-520 (1994)
\item[DS98]~F.Delbaen, W.Schachermayer, The fundamental theorem of asset
pricing for unbounded stochastic processes, {\it Math. Ann.,} 312,
215-250 (1998)
\item[DS01~]~F.Delbaen, W.Schachermayer, Applications to mathematical
finance, In: {\it Handbook of the Geometry of Banach Spaces,}
W.B.Johnson and J.Lindenstrauss, eds., Amsterdam: Elsevier Science
B.V., 2001, pp 367-391
\item[Du93~~]~D.Duffie, {\it Dynamic Asset Pricing Theory,}
Princeton: Princeton University Press, 1993
\item[DH86~]~D.Duffie, C.F.Huang, Multiperiod security markets with
differential information: martingales and resolution times, {\it J. Math.
Econom.,} 15, 283-303 (1986)
\item[EW00~]~J.Eisert, M.Wilkens, Quantum games, {\it J. Mod.
Opt.,} 47, 2543-2555 (2000)
\item[EWL99]~J.Eisert, M.Wilkens, M.Lewenstein, Quantum games and
quantum strategies, {\it Phys. Rev. Lett.,} 83, 3077-3800 (1999)
\item[EK99~]~R.Elliott, P.E.Kopp, {\it Mathematics of Financial
Markets,} New York: Springer-Verlag, 1999
\item[HK79~]~J.M.Harrison, D.M.Kreps, Martingales and arbitrage
in multiperiod securities markets, {\it J. Econ. Theory,} 20,
381-408 (1979)
\item[HP81~]~J.M.Harrison, S.R.Pliska, Martingales and stochastic
interfrals in the theory of continuous trading, {\it Stoch. Proc. Appl.,}
11, 215-260 (1981)
\item[Hu00~~]~J.C.Hull, {\it Options, Futures and Other
Derivatives,} 4th Edition, Prentice Hall: Prentice-Hall, Inc.,
2000
\item[KLS87]~I.Karatzas, J.P.Lehoczky, S.E.Shreve, Optimal portfolio and
consumption decisions for a ``small investor" on a finite horizon,
{\it SIAM J. Control and Optimi.,} 25, 1557-1586 (1987)
\item[K96~~~]~D.Kramkov, Optional decomposition of supermartingales
and hedging contingent claims in incomplete security markets,
{\it Probab. Theory Related Fields,} 105, 459-479 (1996)
\item[Kr81~~]~D.M.Kreps, Arbitrage and equilibrium in economies with
infinitely many commodities, {\it J. Math. Econ.,} 8, 15-35 (1981)
\item[Me73~~]~R.C.Merton, Theory of rational option pricing, {\it Bell J.
Econ. Manag. Sci.,} 4, 141-183 (1973)
\item[Me90~~]~R.C.Merton, {\it Continuous-time Finance,} Basil:
Black-Well, 1990
\item[Mr99~~]~D.A.Meyer, Quantum strategies, {\it Phys. Rev.
Lett.,} 82, 1052-1055 (1999)
\item[My93~~]~P.A.Meyer, {\it Quantum Probability for
Probabilists,} Lecture Notes in Math., 1426, Berlin:
Springer-Verlag, 1993
\item[vN55~~]~J.von Neumann, {\it Mathematical Foundations of
Quantum Mechanics,} Princeton: Princeton University Press, 1955
\item[P92~~~~]~K.R.Parthasarathy, {\it An Introduction to Quantum
Stochastic Calculus,} Basel: Birkh\"{a}user Verlag 1992
\item[PX97~]~G.Pisier, Q.Xu, Non-commutative martingale
inequalities, {\it Comm. Math. Physics,} 189, 667-698 (1997)
\item[S65~~~~]~P.A.Samuelson, Proof that properly anticipated prices
fluctuate randomly, {\it Industrial Manage. Review,} 6, 42-49 (1965)
\item[Sc01~~]~W.Schachermayer, Introduction to the Mathematics of
Financial Markets, to appear in {\it Lecture Notes in Math.,} (St.
Flour summer school 2000)
\item[Sch66~]~H.H.Sch\"{a}fer, {\it Topological Vector Spaces,} Berlin:
Springer-Verlag 1966
\item[Sh99~~]~A.N.Shiryaev, {\it Essentials of Stochastic Finance $(Facts,
Models, Theory),$} Singapore: World Scientific, 1999
\item[St90~~]~C.Stricker, Arbitrage et lois de martingale, {\it Ann. Inst.
H. Poincar\'{e}, Probab. Statist.,} 26, 451-460 (1990)
\item[Ta72~~]~M.Takesaki, Conditional expectations in von Neumann
algebras, {\it J. Funct. Anal.,} 9, 306-321 (1972)
\item[Ta01~~]~M.Takesaki, {\it Theory of Operator Algebras I,}
New York: Springer, 2001
\end{description}
\end{document}